# Integrating optimization with thermodynamics and plant physiology for crop ideotype design

Talukder Z. Jubery[1], Baskar Ganapathysubramanian[1], Matthew E. Gilbert[2], and Daniel Attinger[1]

[1]Department of Mechanical Engineering and Department of Electrical and Computer Engineering, Iowa State University, Ames, IA, 50011;

[2]Department of Plant Sciences, University of California, Davis, CA 95616, USA.

**\***Correspondence to: *baskarg@iastate.edu, gilbert@ucdavis.edu, attinger@iastate.edu*

## Abstract

A computational framework integrating optimization algorithms, parallel computing and plant physiology was developed to explore crop ideotype design. The backbone of the framework is a plant physiology model that accurately tracks water use (i.e. a plant hydraulic model) coupled with mass transport ($CO_2$ exchange and transport), energy conversion (leaf temperature due to radiation, convection and mass transfer) and photosynthetic biochemistry of an adult maize plant. For a given trait configuration, soil parameters and hourly weather data, the model computes water use and photosynthetic output over the life of an adult maize plant. We coupled this validated model with a parallel, meta-heuristic optimization algorithm, specifically a genetic algorithm (GA), to identify trait sets (ideotypes) that resulted in desired water use behavior of the adult maize plant. We detail features of the model as well as the implementation details of the coupling with the optimization framework and deployment on high performance computing platforms. We illustrate a representative result of this framework by identifying maize ideotypes with optimized photosynthetic yields using weather and soil conditions corresponding to Davis, CA. Finally, we show how the framework can be used to identify broad ideotype trends that can inform breeding efforts. The developed presented tool has the potential to inform the development of future climate-resilient crops.

## Introduction

To ensure food security, crop grain yields should be increased globally by 70–100% within the next 40 years [1]. To increase yields, plant breeders and plant scientists are working to develop improved and appropriate varieties of crops. However, the intrinsic uncertainty of climate change, limited water supply and reduction of agricultural land increase the challenges in the crop development process [2]. We have limited time and resources to select the most appropriate crop varieties, and crop modeling provides a rational approach to designing new crop varieties [3].

Traditional methods for finding the best crop varieties, or ideotypes [4], rely on agronomic experiments. The evaluated ideotypes are restricted in time and space, making results site- and season-specific, and the experiments are time consuming and expensive. The use of crop models has greatly enabled crop breeding by reducing costs and accelerating the process of identifying/designing ideotypes.

Many physiological models of crops [5]–[15] have been developed since the pioneering 1948 model of van den Honert [16]. These models are focused on specific aspects of plant physiology: *water transport, time-dependence, influence of environmental conditions, heat and mass transfer, effect of plant geometry, nutrient transport, plant growth, or phloem transport*. While these models can determine and explain optimum relationships between existing traits [5], there is increasing interest in coupling them with optimization tools to identify the most promising traits for a desired response [24], [25]. Over the last three decades, publications on numerical optimization methods have emerged and their number has grown at a rate higher than the growth rate of publications on traditional plant breeding. Considering also that breeding is, per se, an optimization problem, the recent emergence of limited publications at the intersection of "breeding" and "numerical optimization" is not unexpected. Some of these publications describe numerical methods inspired by animal breeding strategies [17], [18], while others seek to optimize the management of a breeding program [19], [20], either by improving the phenotyping associated with breeding [21], or minimizing the genotyping efforts [22]. While these models can determine and explain optimum relationships between existing traits [5], [10], [23], there is work yet to be done to efficiently leverage this knowledge to direct breeding efforts. This is the motivation of our work. We describe our model and how we integrate it with an optimization framework. We then demonstrate an application of the framework by identifying crop ideotypes for a specific location parameterized by weather and soil conditions.

## Materials and Methods

**Plant physiology model**
The backbone of our framework is a mechanistic crop physiological model that is based upon a detailed one-dimensional representation of plant hydraulic characteristics. Liquid-phase, plant-water relations are simply represented as a static series of conductances resistances for stems, leaves and roots ( disregarding capacitive behavior, i.e. stems, leaves and roots do not store any water) , as described in the seminal work of van den Honert [16]. The model is a physiologically explicit representation of C4 maize water-use after canopy closure. The model explicitly accounts for energy balance (convection, radiation, latent heat), transpiration, intercellular $CO_2$ concentration (via both diffusion and biochemical processes), weather conditions (temperature, precipitation, pressure, radiation), and soil type. Seven plant hydraulic 'traits' are considered within the model, as shown in Fig. 1, and can be used to represent the response of leaf evapotranspiration to environmental variation. We next provide details of each submodel that is used to construct the full plant model. Figure 2 shows the various submodels schematically.

*Water transport submodel*

Uptake of water by the root from the soil reaches the top of the canopy due to cohesive-adhesion interactions, but the main driving force for this transport is the dryness of the atmosphere. Water travels from the soil to the leaf as a liquid. Subsequently, as a gas it evaporates from the leaf (through the stomatal pores) to the surrounding environment. The evaporation rate depends on the leaf temperature, external relative humidity, air temperature and boundary layer effects. This is called environmental water demand. To fulfill this demand, the plant supplies liquid water to the leaf. This flow is driven by the potential difference of water between the soil and the leaf and is controlled by the hydraulic resistance of the plant.

Using a one-dimensional representation of plant hydraulic characteristics, as shown in Fig. 1, the water supply, $J_{W,s}$, can be expressed as [26]

$$J_{W,s} = \left( \frac{1}{K_{plant}} + \frac{1}{K_{soil}} \right)^{-1} (\psi_{soil} - \psi_{leaf}), \tag{1}$$

where $K_{plant}$ is hydraulic conductance of the plant, $K_{soil}$ is hydraulic conductance of soil; $\psi_{leaf}$ is the water potential at the leaf, $\psi_{soil}$ is water potential at the soil. Water potential is a combined effect of hydrostatic pressure, osmotic pressure, matric pressure, and gravitational pull.[1]

The hydraulic conductance of the plant can be expressed as

$$K_{plant} = \left( \frac{1}{K_{root}} + \frac{1}{K_{stem}} + \frac{1}{K_{leaf}} \right)^{-1}, \tag{2}$$

where $K_{root}$ is the hydraulic conductance of the root, $K_{stem}$ is hydraulic conductance of the stem, and $K_{leaf}$ is hydraulic conductance of the leaf.

The hydraulic conductance of the soil, $K_{soil}$, depends on the type of soil, the amount of water in the soil and the relative occupancy of the root in the soil. The effect of these parameters is captured via the following equation [27],

$$K_{soil} = k_{sat} \left( \frac{\psi_{sat}}{-\sqrt{\psi_{soil}\psi_{soil,root}}} \right)^{2+3/b} \frac{2\pi L H_s}{\log\left( \frac{\sqrt{\pi L}}{r_{root}} \right)}, \tag{3}$$

---

[1]. Osmotic pressure that depends on the presence of ions in the water is neglected in our model, as we considered the water as pure and free from any minerals.

where $k_{sat}$, b, $\psi_{sat}$ and $\psi_{soil}$ vary among the types of soil, and they represent saturated hydraulic conductivity, texture, water potential of saturated soil, and water potential of the soil, respectively. The rest of the terms are used to capture the effect of the presence of root on the soil conductance. The symbols L, $H_s$ and $r_{root}$ represent root length density of the absorbing root (length per soil volume), depth of the soil occupied by the root and radius of the root.

The water potential of the soil, $\psi_{soil}$, can be expressed as a function of soil water content using an empirical equation developed by Campbell et al. [28] as

$$\psi_{soil} = \psi_{sat}\left(\frac{\theta_{sat}}{\theta}\right)^b, \tag{4}$$

where $\theta_{sat}$ is the saturated water content in the soil, and $\theta$ is current volumetric water content in the soil. In this model, soil water content would gradually deplete as plants fulfill the atmospheric water demand. The depletion of water due to evaporation of water from the soil was not considered here. There is no addition of water in the case of drought conditions. However, under irrigated conditions, based on the irrigation frequency, water is added to the soil until water content reaches $\theta_{sat}$ of the soil. e.g., for irrigation frequency 7, the soil is fully saturated every 7*24 hours.

Soil water potential at the root (Eq. 3), $\psi_{soil,root}$, can be evaluated from leaf water potential and water demand by the plants,

$$\psi_{soil,root} = \psi_{leaf} + J_{W,d}\left(\frac{1}{K_{plant}} + \frac{1}{K_{soil}}\right), \tag{5}$$

Water demand is driven by the gradient of water vapor concentration between the leaves and the surrounding environment and is controlled by the stomatal conductance and air boundary layer conductance. It can be expressed as [26]

$$J_{W,d} = \left(\frac{1}{g_{st}} + \frac{1}{g_{blc}}\right)^{-1}\left(\frac{P_{vl} - P_{va}}{P_a}\right), \tag{6}$$

where $P_{vl}$ and $P_{va}$ represent water vapor pressure in the leaf and atmosphere, respectively. Water vapor pressures are evaluated using Tetens formula [29], $P_{v,i} = RH_i c_0 e^{\left(\frac{c_1 T_i}{T_i + c_2}\right)}$, where RH is the relative humidity, $c_0$ = 0.617 kPa, $c_1$ =17.38, and $c_2$= 239°C. Generally, the leaf inter-cellular space is close to equilibrium with the cells having a relative humidity of greater than 99%, and thus for each of calculation of evaporation we consider the leaf to be fully saturated. $g_{st}$ and $g_{blc}$ are the stomatal conductance and boundary layer conductance to the water vapor transport, respectively.

Boundary layer conductance to water vapor, $g_{blc}$ depends on the atmospheric wind speed and the morphology as well as the orientation of the leaf. Wind speed and leaf dimension are designated as $U_c$, and d as in [28]. Conductance of water vapor through the air boundary layer on the leaf can be considered as forced convection and can be expressed via an empirical equation. Note that, here, contribution from the free convection is neglected, as the ratio of dimensionless parameters $Re^2/Gr$ which reflects the forced convection/free convection is usually much greater thanone. The empirical correlation among the dimensionless Reynolds number, Re, and Schmidt number, Sc, and the conductance can be calculated as,

$$g_{blc} = \alpha \frac{D_{WV} Re^{1/2} Sc^{1/3}}{d_e}, \quad (7)$$

where $Re = \frac{U_c d_e}{\nu_a}$; $Sc = \frac{\nu_a}{D_{WV}}$; α=0.644*1.4 is an empirical parameter; and $d_e$ =0.72d, with d being the width of the maize leaf and 0.72 being used to find the equivalent parabola of the leaf where wind is flowing in the width direction of the parabola. $U_c$, $\nu_a$ and $D_{wv}$ represent the wind speed on the top of the canopy, kinetic viscosity of air and water vapor diffusivity in air.

Wind speed can increase approximately logarithmically with distance above a plant canopy, and is also influenced by the plants. The variation in wind speed can be described by

$$U_c = \frac{U^*}{0.4} \ln \frac{H_c - mH_c}{nH_c}, \quad (8)$$

where 0.4 is related to the von Karman constant, $H_c$ is the height of the plant, $mH_c$ is the zero-plane displacement, and $nH_c$ is the roughness length. Generally, m is 0.7 and n is 0.1. $U^*$ is termed the shearing or friction velocity and can be calculated from the wind speed $U_m$ that is measured at height $H_m$ from the ground as

$$U^* = \frac{0.4 U_m}{\ln \frac{H_m - mH_c}{nH_c}}. \quad (9)$$

Only around 3% of water that is absorbed from the soil is used by the plant for metabolism/growth, and less than 0.1% is used for photosynthesis.

*$CO_2$ transport and net photosynthesis submodel*

Along with water, the plant needs $CO_2$, sunlight and enzymes for photosynthesis. From the environment, gaseous $CO_2$ diffuses into the leaf via stomata and then dissolves in water and diffuses to the cells where photosynthesis takes place. The consumption of $CO_2$ during photosynthesis depends on the sunlight and enzyme activity (plants always have sufficient water to split in photosynthesis).

The rate of gaseous $CO_2$ transport to the leaf is named as $CO_2$ supply. The supply is driven by the $CO_2$ concentration gradient between the atmosphere and the leaf inter-cellular space, and is controlled by the conductance of stomata and the air boundary layer. This supply can be expressed as [26]

$$J_{C,s} = \left(\frac{\beta}{g_{st}} + \frac{\chi}{g_{blc}}\right)^{-1} (C_{C,a} - C_{C,i}), \tag{10}$$

where β and χ are the ratio of $CO_2$ conductance and water vapor conductance through stomata and air boundary layer, respectively. β is the ratio of the molecular diffusivities of $H_2O$ and $CO_2$, χ is power ¾ of β, and $C_{C,a}$ and $C_{C,i}$ are the concentration of $CO_2$ at the atmosphere and inside the intercellular space of the leaf.

The demand of atmospheric $CO_2$ depends on the supply of sunlight and the performance of the enzymes that control photosynthetic activity. The plant gets some $CO_2$ as a byproduct of metabolism or respiration activity in the mitochondria and lowers the atmospheric $CO_2$ demand.

For C4 plants, the electron transport to support $CO_2$ reduction occurs in mesophyll (C4 cycle) and bundle-sheath (C3 cycle) cells. If the supply of sunlight is lowered compared with enzyme performance, which mainly occurs during the morning, sunset, or cloudy days, the photosynthetic rate can be expressed as [30]

$$J_{C,d}(light) = \frac{(1-x)J_{e,t}}{3} - R_t, \tag{11}$$

where $J_{e,t}$ is the total electron transport rate is at leaf temperature, $R_t$ is the rate of $CO_2$ production from respiration in the mesophyll and bundle sheath cell, and $x$ is a fraction of total electrons that are used by the mesophyll.

PEPCase, phosphoenolpyruvate carboxylase, and Rubisco are two enzymes that significantly control the photosynthesis activity in C4 plants. PEP (three-carbon backbone) controls the activity of the mesophyll cell (it catalyzes the primary carboxylation in a tissue that is close to the external atmosphere) and Rubisco controls activity in bundle sheath cell. In the case of no limitations on the supply of reductant to photosynthesis (higher light intensities), the photosynthetic demand can be expressed as, [30]

$$J_{C,d}(enzyme) = \min \begin{cases} (V_{PEP} + g_{C,bs}C_{C,m} - R_m) \\ V_{RO_{max}} - R_t \end{cases}, \tag{12}$$

Where the top expression in the right-hand side depends on the performance of PEPcase in the mesophyll cell, and the bottom expression depends on the Rubisco performance in the bundle sheath cell. $g_{C,bs}$ is the bundle-sheath conductance to $CO_2$, $C_{C,m}$ is the concentration of $CO_2$ in the mesophyll cell (note that we assume that $C_{C,m} = C_{C,i}$, $CO_2$ concentration in inter-cellular space), $R_m$ is mitochondrial respiration in the mesophyll at leaf temperature (i.e. $CO_2$ supply from the respiration of the mesophyll cell), $R_t$ is the total mitochondrial respiration in the mesophyll and bundle sheath at leaf temperature, and $V_{RO\,max}$ is the maximum rubisco carboxylation rate.

$V_{PEP}$ is the effective PEP carboxylation at leaf temperature. It depends on the availability of $CO_2$ and the regeneration of PEP and can be expressed as [30]

$$V_{PEP} = \min \begin{cases} \dfrac{C_{C,m} V_{PEP_{max}}}{C_{C,m} + K_p}, \\ V_{PEP,R} \end{cases} \quad (13)$$

where the top expression in right-hand side is related to the carboxylation rate of PEP, expressed with the Michaelis-Menten Equation. $C_{C,m}$ is the $CO_2$ partial pressure in Mesophyll, $V_{PEP_{max}}$ is the maximum PEP carboxylation rate at leaf temperature, and $K_p$ is the Michaelis-Menten constant for PEP carboxylase for $CO_2$ at leaf temperature. Note that the Michaelis-Menten constant, $K_p$, refers to the concentration of $CO_2$ at which the reaction rate is half of $V_{PEPmax}$. The carboxylation rate can be decreased if there is not enough PEP, and that depends on the $V_{PEP,R}$, the PEP regeneration rate at leaf temperature.

The temperature-dependent properties in the equations are evaluated using the following equations [31]

$$V_{RO_{max}} = \frac{V_{RO_{max},25} Q_{10,V_{RO_{max}}}^{(T_{leaf}-25)/10}}{\left(1+e^{A(B-T_{leaf})}\right)\left(1+e^{A(T_{leaf}-C)}\right)}, \quad (14)$$

$$V_{PEP_{max}} = V_{PEP_{max},25} Q_{10,V_{PEP_{max}}}^{(T_{leaf}-25)/10}, \quad (15)$$

$$V_{PEP,R} = V_{PEP,R_{25}} Q_{10,V_{PEP,R}}^{(T_{leaf}-25)/10}, \quad (16)$$

$$V_{PEP} = V_{PEP,25} Q_{10,V_{PEP}}^{(T_{leaf}-25)/10}, \quad (17)$$

$$R_m = R_{m,25} Q_{10,R_m}^{(T_{leaf}-25)/10}, \quad (18)$$

$$R_t = R_{t,25} Q_{10,R_t}^{(T_{leaf}-25)/10}, \quad (19)$$

$$J_{e,t} = J_{e,25} e^{J_a - \dfrac{J_b}{0.00831(273.15+T_{leaf})}}, \quad (20)$$

where, A, B, C, $J_a$ and $J_b$ are physiological parameters related to the carboxylation rate and electron transport rate. The subscript 25 in the symbols indicates the parameters at 25°C. Hourly $J_{e,25}$ can be expressed as [30]

$$J_{e,25} = \frac{I + J_{e_{max},25} - \sqrt{(I + J_{e_{max},25})^2 - 4\lambda I J_{e_{max},25}}}{2\lambda}, \tag{21}$$

where $\lambda$ is the empirical curvature factor and $I = PAR \times f_{PAR\_PSII}$. $f_{PAR\_PSII}$ is the fraction of PAR that contributes to the Photosystem II.

Using the photosynthesis rate of the above two limiting cases, the $CO_2$ demand can be expressed as [30]

$$J_{C,d} = \min(J_{C,d}(light), J_{C,d}(enzyme)), \tag{22}$$

*Energy balance on leaf submodel*
In the above equations many of the parameters related to leaves, for instance, water vapor pressure, enzyme activities, etc., depend on the leaf temperature. Leaf temperature can be evaluated by using first principles in so-called "big leaf models" [28]. Several assumptions are considered in this model: the leaf is flat and perpendicular to the incident sunlight; leaf does not store any energy; and energy storage ; and there is negligible heat generation due to metabolic activity in the leaf.. Considering the leaf is at steady state, the energy balance equation on a leaf can be expressed as [28]

$$[a(1+r)S + a_{IR}\sigma(T_a)^4] - [2e_{IR}(T_{leaf})^4 + C_p g_{hbc}(T_{leaf} - T_a) + J_{W,d} L_{vap}] = 0, \tag{23}$$

where the terms are energy input by solar irradiation and the surrounding irradiation, cooling by leaf irradiation, convective/conductive cooling by the air/temperature gradient and heat loss accompanying water evaporation. In Equation (23), a is the absorptance of the leaf, r is the reflectance, S is the solar irradiation, $a_{IR}$ is the absorptance of leaf for thermal infrared radiation, $L_{vap}$ is the latent heat of vaporization of water, $h_c$ is the convective heat transfer coefficient, and $g_{hbc}$ is the air boundary conductance to heat transfer.

The boundary layer conductance depends on leaf morphology and wind speed, and can be expressed via empirical relationships of dimensionless parameters Reynolds number, Re, and Prandtl number, Pr. It can be expressed as

$$g_{hbc} = \beta \frac{D_H \text{Re}^{1/2} \text{Pr}^{1/3}}{d_e}, \tag{24}$$

where $\text{Re} = \dfrac{U_c d_e}{v_a}$; $\text{Pr} = \dfrac{v_a}{D_H}$; β=0.644*1.4 is an empirical parameter; $d_e$ =0.72d, with d being the width of maize leaf and 0.72 being used to find the equivalent parabola of the leaf where the wind is flowing in the width direction of the parabola. $U_c$, $v_a$ and $D_H$ represent the wind speed on the top of the canopy, kinetic viscosity of air and thermal diffusivity in air. The effect of the temporal variation of soil is not explicitly included in Equation (23). Instead, the effect was implemented using the FAO56 algorithm, as in [32].

*Stomatal conductance submodel*
In the pathway of the supply of $CO_2$ (Eq. (10)) from the environment and demand of $H_2O$ (Eq. (6)) to the environment, stomatal conductance is the most significant parameter. In general, stomatal conductance is around several orders of magnitude lower than that for air boundary layer conductance. Stomatal conductance is a very complex parameter that is affected by environment, plant physiology and heredity.

At least 35 empirical models have been proposed to capture the complex relationship between stomata conductance and various factors including [5], [8], [33]–[38] Such factors include environmental factors, for example, solar radiation, soil water content, humidity and wind speed, etc., and physiological factors, for example, leaf water potential, root water potential, hydraulic root conductance, etc. Few models explicitly include the plant physiological influences on the stomatal conductance apart from entirely empirical functions. Here, we propose a model which is developed based on the sigmodal response of the stomatal conductance with respect to the leaf water potential [39]. The main concept of this model is shown in Fig. 1. Here, the stomatal conductance will start decreasing when leaf water potential touches the threshold potential, which depends on the plant genotype. Closing rate is controlled by the two sensitivity terms $S_l$ and $S_r$ and also the root water potential. The model is expressed as,

$$g_{st} = g_{min} + \dfrac{\min\left(g_{max}, g_1 (J_{C,d})^{g_2}\right)}{1 + 0.0526 \left(\dfrac{\psi_{leaf}}{\psi_{th}}\right)^{\frac{(S_l - \psi_{root} S_r)}{Z}}}, \quad (25)$$

where the environmental response on the stomatal conductance is implicitly influenced by $J_{C,d}$ and $\psi_{leaf}$. $\psi_{th}$ is the threshold bulk leaf water potential at stomatal closure, and $S_r$ is the slope of the relationship between stomatal conductance and root water potential, $\psi_{root}$. $g_1$ and $g_2$ are plant physiological properties related to photosynthesis. Z is a parameter to make the exponent dimensionless.

*Method to evaluate net photosynthesis and water usage*
Figure 2 shows the schematic of the concept and Fig. 3 shows the flow chart of the model implementation. For the input weather condition, soil and agronomic/management practices the net photosynthesis and water transpiration (Tr) can be evaluated iteratively by satisfying (Eq. 6) (Eq. 10), (Eq. 23) and (Eq. 25). A plant is considered dead and net photosynthesis is zero if the plant experiences a permanent wilting condition or permanent temperature damage. Both states cause irreversible damage to the plant.

*Framework for crop design*

Ideotype design requires identifying the optimal combination of plant physiological traits to maximize photosynthesis for specific environmental conditions and management practices. We formulate the design problem as an optimization problem. Thus, by writing photosynthesis as the following functional form,

$$AN = f(\text{Traits})$$

The optimization problem is defined as

$$\underset{(\text{Traits}=\text{Trait}1,\text{Trait}2....)}{\arg\max} \sum_{i=1}^{N} \alpha_i AN_i \,, \tag{26}$$

where i represents different conditions related with agronomic practices (e.g. no-irrigation, weekly irrigation, etc.), weather conditions, or soil type. $\alpha$ is a weighing factor that depends on the preference of the designer.

*Physiological traits and location/weather/management conditions*

*Physiological Traits:* In our model, a plant has been represented by 37 physiological traits. Typical values of most of those traits were collected from the current literature (Table 1). Note that these traits represent the adult crop. The traits used in the photosynthetic submodel were collected via gas exchange calibration.

Among the 37 traits in this study, we considered seven hydraulic traits: minimum stomatal opening ($g_{min}$); maximum stomatal opening ($g_{max}$); sensitivity of stomatal opening with leaf water potential ($S_l$); threshold bulk leaf water potential at stomatal closure ($\psi_{th}$); sensitivity of $S_l$ with root water potential ($S_r$); shoot hydraulic conductance ($K_{shoot}$); and root hydraulic conductance ($K_{root}$). These traits affect the stomatal conductance which is a vital trait for photosynthesis [ref]. For the optimization problem, those traits were bounded within the ranges in Table 2, ranges currently found in nature.

*Traits related to the photosynthetic submodel:* Parameters used in the photosynthetic submodel are difficult to find in literature. Seventeen physiological parameters used in the model (that are related to the photosynthesis equations 14-20) were calibrated using gas exchange data. The net photosynthetic rate ($A_n$) was calibrated using gas exchange measurements made on leaves of two maize plants grown in mini-lysimeters at the Davis Agricultural experimental station in June to July 2013. Net photosynthetic rate was modelled as a function of three inputs: intercellular $CO_2$ concentration, photosynthetically active radiation (PAR) and leaf temperature. Thus, these three variables were varied using a LI-COR 6400 gas exchange system to obtain sufficient variation to calibrate the photosynthesis submodel.

The $CO_2$ response data is shown in Fig. 4(a-b) and the PAR-light-response data in Fig. 4(c-d). The entire dataset was used to calibrate the C4 photosynthetic parameters using an optimization algorithm.

*Location/weather/management conditions:* We considered a drought-prone environment condition, i.e. Davis, CA in 2010 June-July (see S1) with clay soil (see Table 1) and irrigation frequency of seven days.

### *Method to implement crop design framework*

There are several approaches to solve this optimization problem. Here, we utilize a gradient free, evolutionary optimization strategy. This strategy is selected because, as Figure 5 shows, the cost functional (Equation 26, when varying only two traits) is non-convex and corrugated. This highly-corrugated surface has many local maximum. This precludes the utilization of gradient-based methods, and instead suggests the applicability of stochastic, multistart methods that can explore the phase space efficiently. We specifically use a genetic algorithm (GA) (a gradient-free meta-heuristic evolutionary search algorithm) to identify the optimal traits. GA is well suited to multi-modal, highly corrugated solution spaces, especially when the cost function is not easily adapted to gradient-based methods.

Because GA deploys a population of potential solutions distributed over the design space, they are less prone to getting stuck in shallow local minima. GA is an inherently stochastic method, so we repeat each optimization multiple times (10 times) to consider statistical significance of results and attempt to reliably explore the phase space. The implementation framework can be found in Figure 6.

## Results and discussion

### Plant Physiology Model Validation

The physiological model is implemented in MATLAB with inputs of soil and hourly weather data over a 60-day period. Each model evaluation for a given trait configuration – producing hourly outputs – took about 40 seconds on a standard laptop.

The plant physiology model builds on the water transport model, and the temperature model depends on the conservation principles, which are inherently satisfied in our method. Therefore, we perform a validation exercise on the photosynthetic submodel. The excerise was performed where marked leaves on 14 maize plants, growing adjacent to the calibration plants, were monitored repeatedly using the LI-COR 6400 gas exchange system for a day. During that period the plants were subjected to a diurnal gradient of low to high ambient temperature, and a range of light. A subset of the plants also had water withheld to evaluate the photosynthetic submodel's performance under water stress.

The photosynthesis submodel, trained on the light, temperature and $CO_2$ response curves, successfully predicted the photosynthesis of the 14 validation plants during the day of drought and varying temperature (Fig. 7). Current models of photosynthesis do not account for major damage to the photosynthetic apparatus in a mechanistic manner. Thus, the model is unable to predict photosynthetic rates of a couple of points that represented very severely stressed plants.

**Design of ideotypes**

We deployed the crop design framework on the computing clusters available at Iowa State (CyEnce cluster) and via NSF XSEDE resources at TACC (Stampede). The simulations usually took about 4 hours to run for each optimization run on a server with 16 core 2.0 GHz Processer with 128 GB RAM. Optimizations were initialized with different random seeds and rerun 10 times. In this process, over one million distinct trait combinations were evaluated, and 10 ideotypes were designed.

**Comparison between Designed Crop and a Typical Crop**

Figure 8 (A) shows that the ideotype produces 10% higher net photosynthesis (yield) than that of the typical maize. To investigate that, we compare and explore the performances of those two crops on the hottest day of the season, June 27. The weather on that day is shown in Figure 8, with an average daytime temperature of $39.81^0C$. Starting from the morning, the hourly value of solar radiation increases till midday and then decreases till sunset. Relative humidity is high at night and it decreases during the daytime. Hourly precipitations on that day are zero. It is noted that the atmospheric temperature increased as the day progresses and went as high as $43^0C$. This temperature is higher than the optimum functional temperature of maize plants. Thus, plants that can cool their leaves are desirable.

Figure 8 (D-E) reveals that for the above weather inputs, at the early and later part of the day, there is no significant variation of hourly net Photosynthesis (An) between the two plants. However, significant variation is observed at the midday. Photosynthesis depends on CO2, PAR, enzyme and temperature.

Midday generally has enough PAR, so An depends on CO2 supply, enzyme performance and temperature. In our study, the enzyme performance profile is the same for both of the plants. Therefore, midday variation of An depends on the supply of CO2 and temperature.

Figure 8 shows that during the midday period the average CO2 supply is 180 and 200 ppm for the typical maize and ideotype, respectively. The leaf temperature of the ideotype is lower than that of the typical maize. These two conditions enable a higher An for the ideotype (Fig 8). The lower temperature facilitates a shorter duration in which the leaf temperature is above the optimum temperature for photosynthesis.

Reduced CO2 and lower temperature are related to high stomatal conductance, as shown in Fig. 8. Therefore, Figure 8 (F &G) shows that the typical maize plant has lower stomatal conductance than that of an ideotype, but should create higher concentration gradient by lower $CO_2$ concentration than that of an ideotype in the early part of the day. Figure 8 (E & H) shows that, due to low stomatal conductance of the typical plant, the cooling of the leaf due to transpiration of water is lower, and as a result, leaf temperature is higher than that of an ideotype. In short, the main driver behind the increase of photosynthesis for the ideotype is the positive shift of stomatal opening operating range.

**Values and Significance of the Traits of the optimized ideotype**
The positive shift of stomatal conductance depends on the convoluted effect of the seven traits. Here, we present and discuss on the values and significance of the traits.

*Designed Values of the Traits*
Our framework provides ten different combinations of traits for the designed crop (Fig 9), all of them has the same yield (photosynthesis). Based on the variability, we came up with following two hypotheses:
1) any values of traits within the upper and lower values of suggested traits can be a design crop;
2) some traits might be insensitive for our condition and some traits have threshold values after which they are insensitive.
Hypothesis 1 does not hold when we use arbitrary combinations of traits within the ranges (see Fig S2). To test the second hypothesis, we perform sensitivity analysis for all ten combinations by varying one parameter within the allowable range while keeping the others fixed. The sensitivity plots indicate that there are threshold values for all of the parameters. There are higher limits for $S_l$, $\psi_{th}$, $S_r$ and lower limits for $g_{min}$, $g_{max}$, $k_{shoot}$ and $k_{root}$. Maintaining those threshold values and subtly changing the other threshold values in the opposite direction shows the decrease in photosynthesis.

*Significance of the Traits*
In short, photosynthesis mostly depends on the available sunlight, CO2 and temperature. Among these parameters, CO2 and temperature can be optimized by adjusting hydraulic parameters. Among the hydraulic parameters: $g_{min}$ and $g_{max}$ are the most directly constraining. $g_{st}$ (a direct function of $g_{min}$ and $g_{max}$) is the dominating parameter in the transport pathways (causes the highest path resistance for both CO2 and H2O transports). Rate of water transpiration controls the cooling effect on the leaves. Cooling (not freezing) is always beneficial during the night (it reduces the cost of respiration), however, during day it may have positive or negative effect on the photosynthesis based on the optimum temperature for the enzyme activity (Fig 8). Other variables including $k_{shoot}$, Sl, Sr, $\psi_{th}$, $k_{root}$ primarily ensure that the plant does not reach the permanent wilting potential. These values will affect the gst if the plant senses water scarcity when leaf water potential reaches the 'red alert' point, indicated by $\psi_{th}$ value.

More specifically, for our designed ideotype

- $g_{max}$ should be higher than a specific value, so that the plant is able to use its available photosynthetic capacity. The required value of $g_{max}$ mostly depends on the highest solar radiation (PPF: photosensitive photon flux density, mostly 400 and 700 nm).

- $g_{min}$ should be as high as possible to reduce night-time respiration cost, although it may increase the irrigation cost for the season. $g_{min}$ sets the minimum transpiration the plant can do under high VPD or extreme soil water deficit, and thus affects the rate of water depletion under the most extreme of circumstances.

- Our designed ideotypes do not feel water stress (i.e. red alert) in well-watered conditions, and are never forced to adjust gst due to water-related issues. However, an ideotype's $k_{plant}$ and red alert value, $\psi_{th}$, must be selected appropriately. A lower $k_{plant}$ than the specified value may cause a leaf water potential that is lower than the 'red alert' value, i.e. the plant will register water stress. A similar effect will happen if the plant increases (less negative) the $\psi_{th}$ value. Required values for $k_{plant}$ and $\psi_{th}$ have an inverse relationship. Therefore, to reduce the cost of root generation, the plant should operate at the lowest (more negative) possible $\psi_{th}$, i.e. a value close to the permanent wilting potential, thus lowering the $k_{plant}$ requirement, i.e. it refers to less root production.

*Tradeoff between below ground mass and photosynthesis/above ground mass*

Next, we extrapolate our results from our design space (reproductive stage) to vegetative stage. During the vegetative stage, to increase the photosynthetic capability of the plant (LAI), a bigger shoots could be better, whereas smaller roots would help the plant to invest more resources into growing the shoot. Therefore, we explore the effect of a smaller root, i.e. smaller $k_{root}$, on the net photosynthesis of the ideotypes.

Figure 10 shows that a 50% reduction of $k_{root}$ from 40 to 20 reduces the net photosynthesis by only 0.06%. For the reduced $k_{root}$ plant, the leaf water potential sometimes reaches lower than the threshold leaf water potential ($\psi_{th}$) and closes (Figure 11). The modified plant has the potential to increase LAI during the vegetative stage, leading to improvement in net photosynthesis (yield).

## Conclusions

An integrated framework of optimization, thermodynamics and plant physiology was developed to design a crop ideotype. The backbone of the framework is a 1-D plant physiology model and the coupling of transport and energy conversion models based on laws of thermodynamics. The models were augmented with a nature-inspired meta-heuristic optimization method, the genetic algorithm (GA), and was implemented via the MATLAB® software. The framework was used to design maize crop for a drought-prone weather condition in Davis, CA. Seven physiological traits which are primarily related to plant hydraulics and ultimately affect the photosynthesis and water usage were considered in the study. The traits are minimum stomatal opening ($g_{min}$); maximum stomatal opening ($g_{max}$); sensitivity of stomatal opening with leaf water potential ($S_l$); threshold bulk leaf water potential at stomatal closure ($\psi_{th}$); and sensitivity of $S_l$ with root water potential ($S_r$); shoot hydraulic conductance ($K_{shoot}$); and root hydraulic conductance ($K_{root}$). With enough irrigation, the designed crop showed 10% improvement in yield, and $g_{max}$, $g_{min}$ and $\psi_{th}$ are found to be the vital traits. Currently, the model is using hourly data; however, it could be easily modified for more frequent data. The framework is modular and can be easily augmented with other existing mechanistic models to capture more physics. The developed tool can help plant breeders and scientists to determine the optimal crop ideotypes for various climates (climate-smart crops) and locations. Integration of the developed framework with breeding programs can speed the crop development process, wherein the framework can be used to propose ideotypes for target environments and the breeder can breed plants like based on the ideotypes.

Potentially, ideotypes designed using a different crop model might look different from those presented, as shown for wheat in [40]. These models can simulate observed yields under a range of environments for the current conditions. However, simulated climate change impacts could vary across models due to differences in model structures and parameter values [41]. Further improvements of crop models and a more rigorous framework will be required for robust crop ideotype design.

**ACKNOWLEDGEMENTS**

T.Z.J, B.G, D.A gratefully acknowledge financial support from the Presidential initiative for interdisciplinary research of Iowa State University. B.G and T.Z.J gratefully acknowledge the Plant Science Institute at Iowa State University, and computing support via NSF XSEDE CTS110007. M.E.G gratefully acknowledges support via USDA National Institute of Food and Agriculture, Hatch project number #1001480.

**AUTHOR CONTRIBUTIONS:**

D.A formed the interdisciplinary team and proposed to couple numerical optimization and plant physiology, D.A, M.E.G, B.G designed research plan, M.E.G. designed the crop hydraulic model, B.G. implemented model, all authors improved the model, B.G and T.Z.J designed the optimization framework, T.Z.J ran the simulations and processed the data, all authors analyzed data and wrote the paper.

**COMPETING FINANCIAL INTERESTS**

The author(s) declare no competing financial interests.

# Figures and Legends

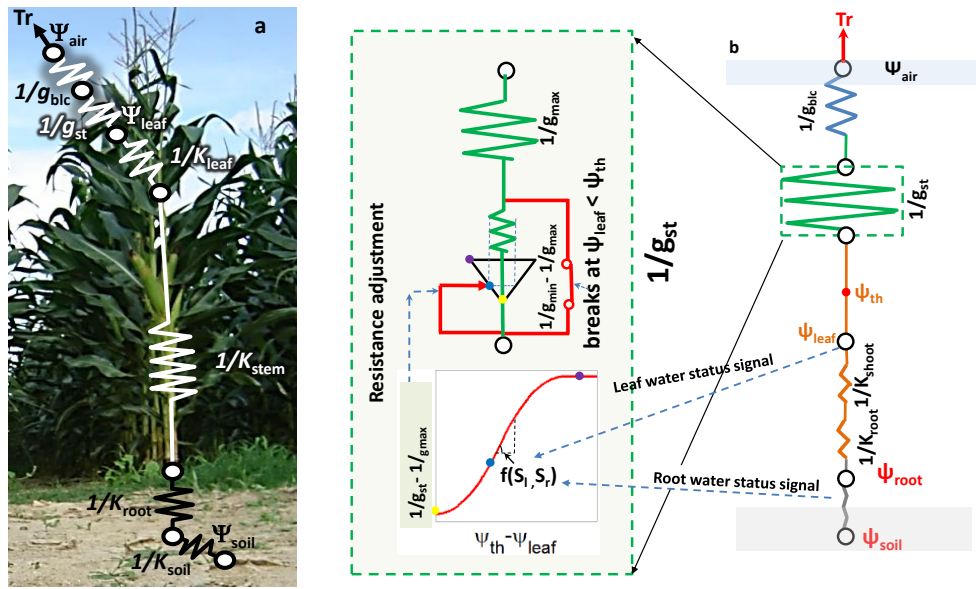

**Figure 1**: A conventional resistance or conductance (resistance=1/conductance) model of maize hydraulics (panel a), and the model used for simulating maize hydraulics including feedbacks (panel b). The conductance of the stomata to water vapor (gst) and CO2 is modulated by the water potential of the leaf ($\psi_{leaf}$) if below a threshold ($\psi_{th}$). The stomatal opening is scaled to the maximum stomatal conductance (gmax; a proxy of how many stomata and how wide they open) which sets the maximum water loss rate and the maximum CO2 uptake rate for sunlit leaves. How effective stomatal closure is (minimum stomatal conductance; gmin) is determined by cuticle waxes which stop water loss from the leaf surface, affecting the rate of desiccation under drought, but this state also prevents CO2 uptake. The slope of the response is tuned by an inherent sensitivity (Sl) or a contribution of the root, based upon sensing of soil drying (Sr). The supply of water is proportional to the difference of water potential between soil and air, and inversely proportional to three conductances in series: Ksoil, Kroot, and Kstem+leaf. The demand for water is driven by environmental variables: the boundary layer conductance and the temperature of the leaf. The leaf temperature is determined through an energy balance and influences both transpiration and a coupled model of photosynthesis.

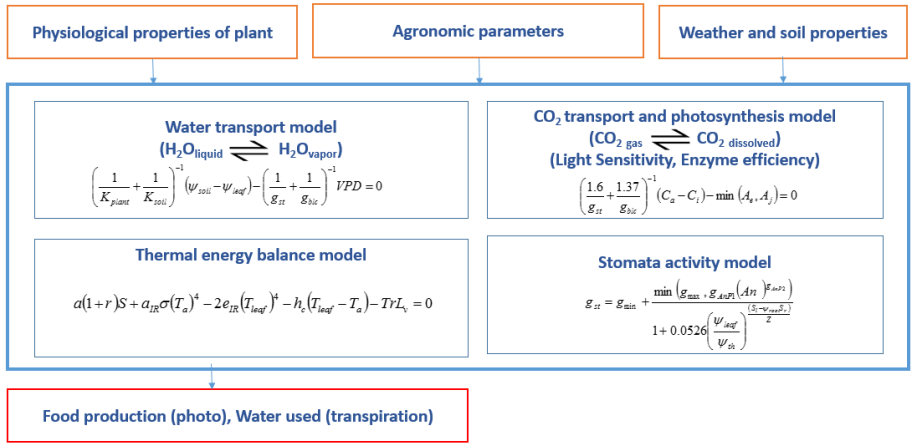

**Figure 2**: Model input, output and connectivity among the submodels.

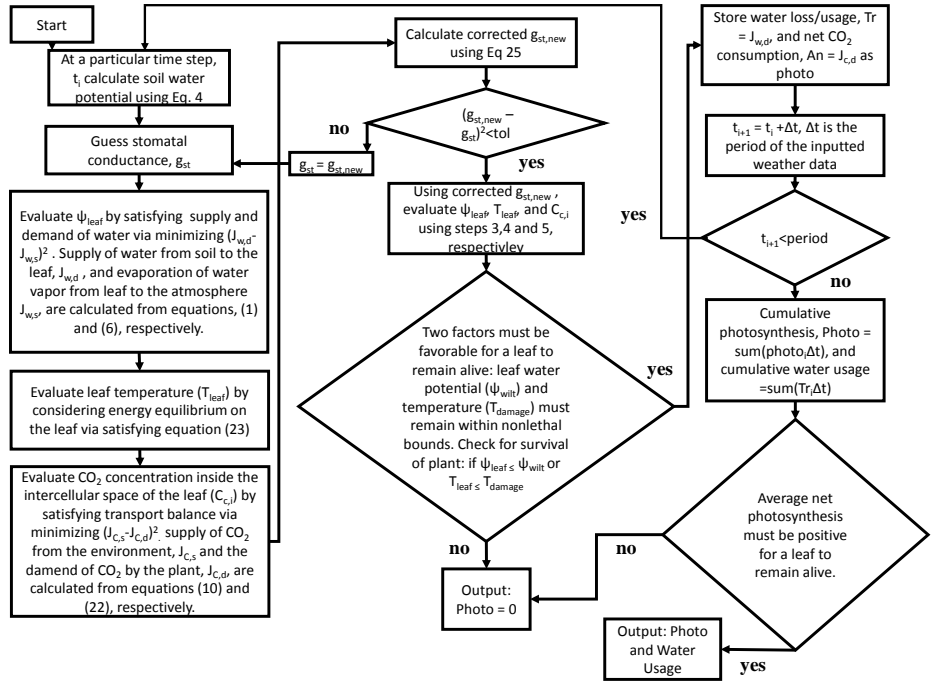

**Figure 3:** Flowchart for the implementation of plant physiology model

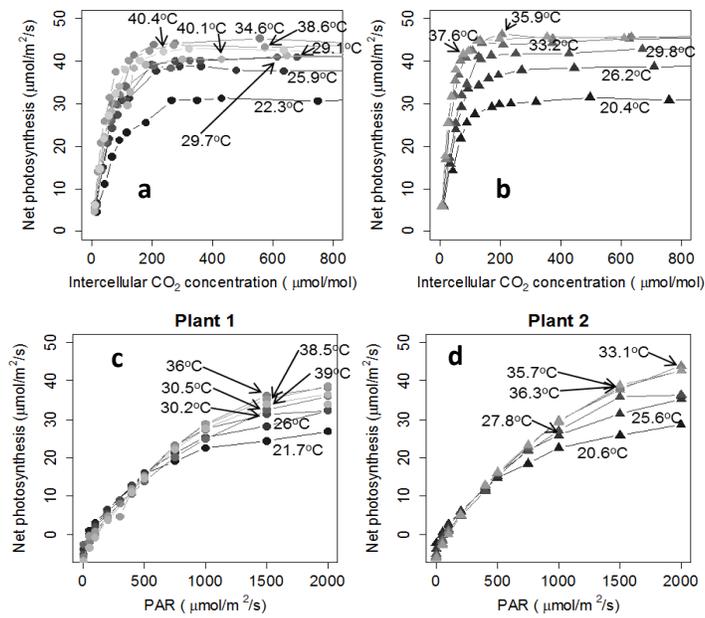

**Figure 4:** (a-b) $CO_2$ responses of maize photosynthesis at varying leaf temperatures for two plants used in the calibration of the photosynthesis submodel. Lines connect points measured at the same leaf temperature. (c-d) Response of maize photosynthesis to photosynthetically active radiation (PAR) measured at varying leaf temperatures for two plants used in the calibration. Lines connect points measured at the same leaf temperature.

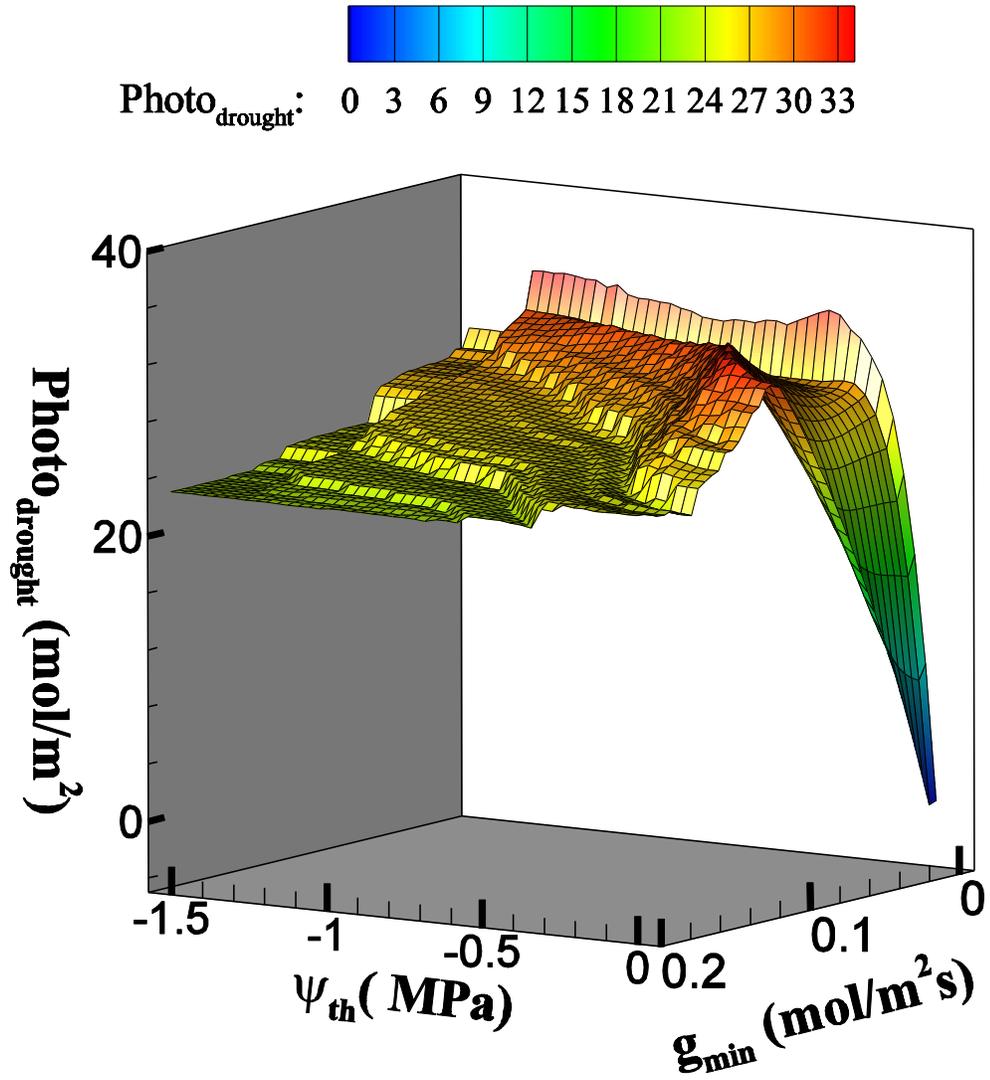

**Figure 5:** Distribution of photosynthesis, in terms of CO2 assimilation, within the selected ranges of gmin and $\psi_{th}$ for typical maize as in Table 1.

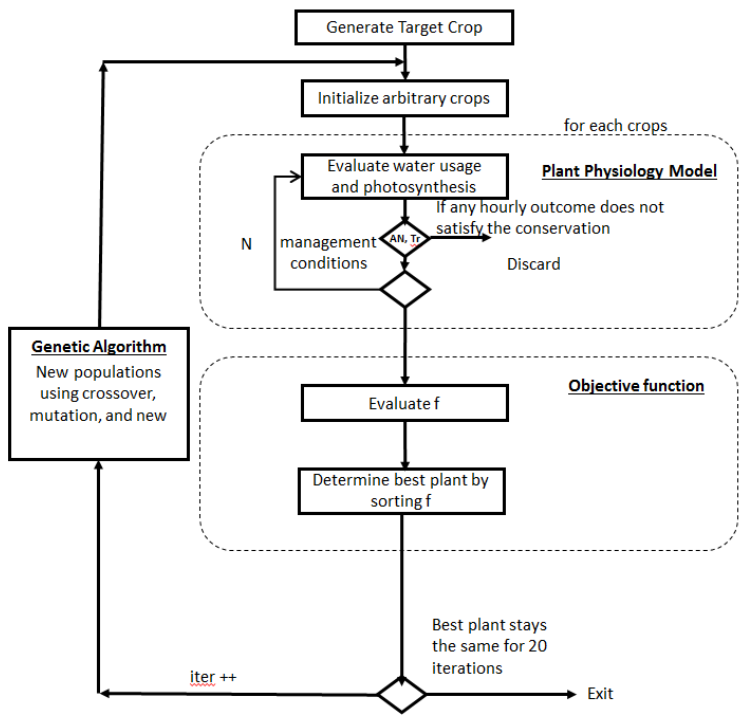

**Figure 6:** Flowchart of crop design framework implementation.

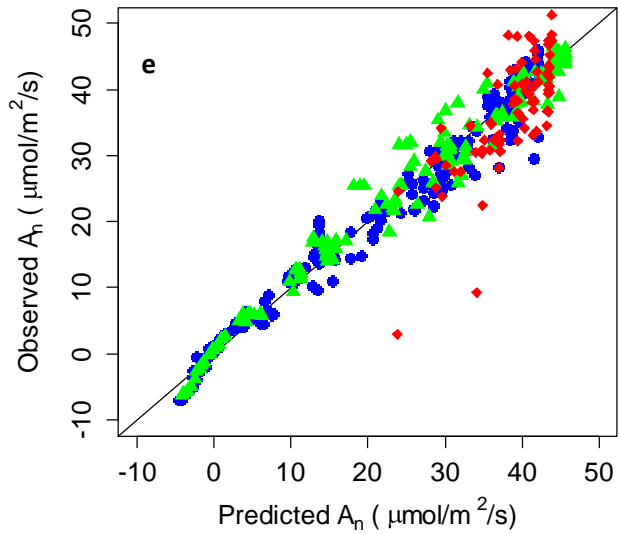

**Figure 7**: Validation of the C4 photosynthesis submodel. The circles represent the predicted and observed calibration data for light, $CO_2$ and temperature responses for plant 1 and triangles for plant 2. The diamonds represent the validation data: the observed and predicted photosynthesis of 14 plants measured over a day varying in drought treatments, temperature and light. The points that deviate from the 1:1 relationship were plants that underwent the greatest drought stress during the hottest time of the day, and represent photosynthetic inhibition or damage.

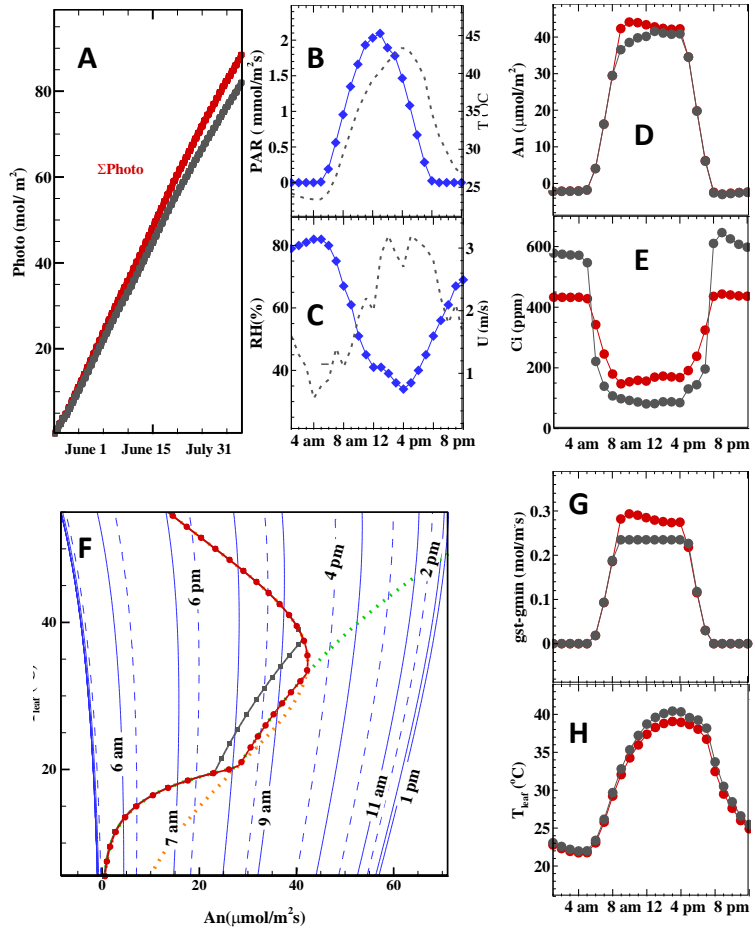

**Figure 8:** (A) Hourly cumulative photosynthesis for the typical plant (red line) and the sets of simulated ideotypes at well-irrigated condition. For simulated plants, the profiles are overlapped. Here, instead of days of the months, time has been presented as hours. (B-C) Hourly variation of solar radiation and relative humidity on June 27. (D-E) Hourly variation of photosynthesis (An), and $CO_2$ concentration inside the leaf. (F) Green for PEP limited case, Orange for Rubisco limited case, Red symbol (for ideotypes) is the net effect with CO2 concentration 180 ppm, Gray symbol (for typical) is the net effect with CO2 concentration 120 ppm. Blue lines are light (PAR) limited case. (G-H) Hourly variation of variation stomatal conductance with respect to minimum conductance, and leaf temperature.

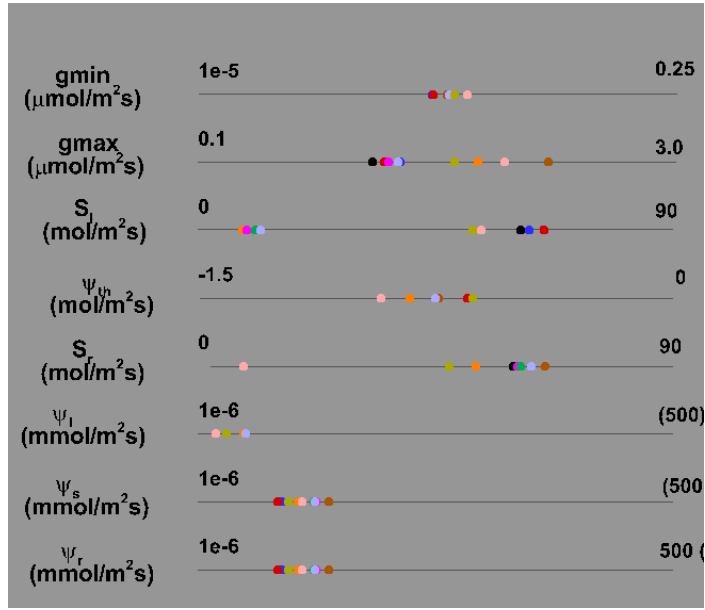

**Figure 9:** Variation of the traits among the outcomes obtained via optimizations. The big red symbol corresponds to the typical value and the numbers indicate the upper and lower limits of the traits (Table 1).

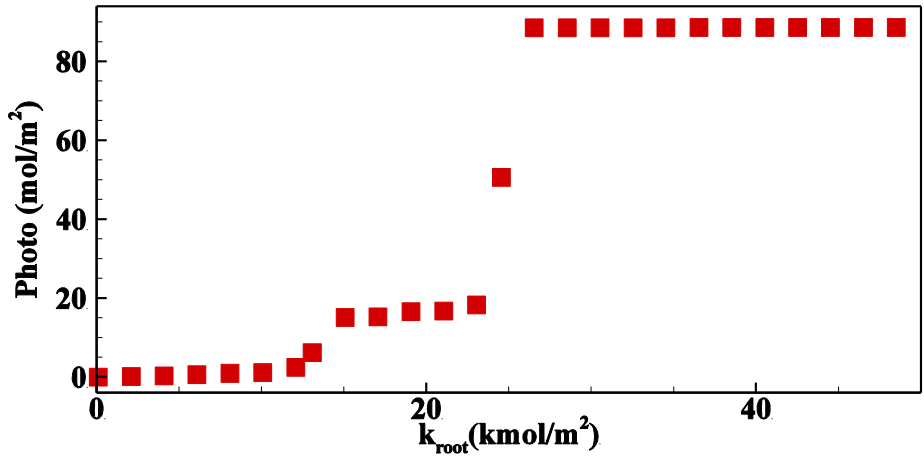

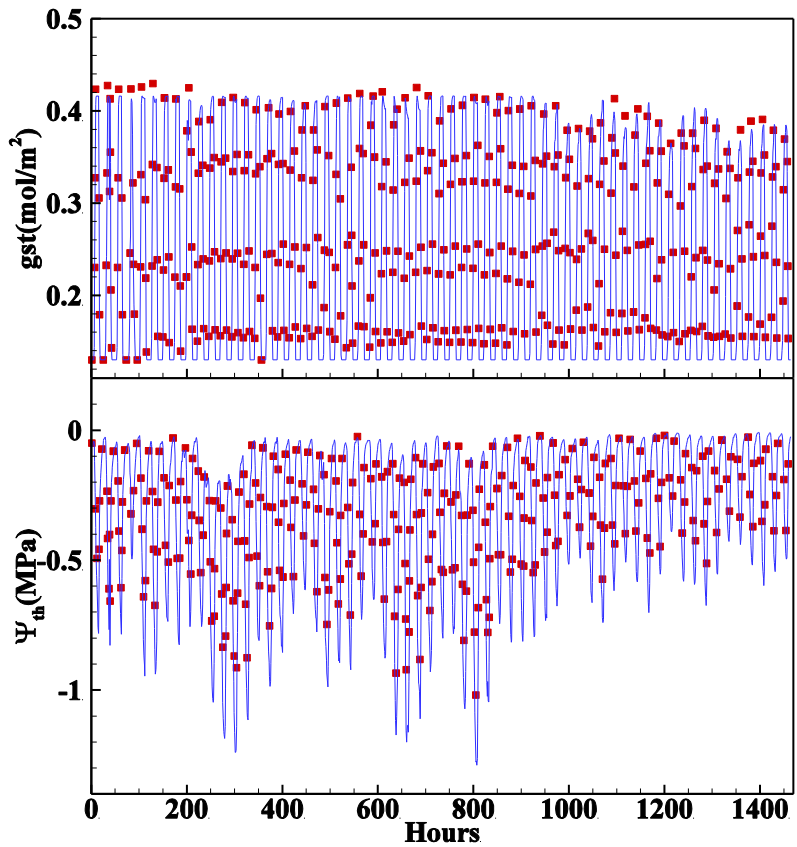

**Figure 10:** (top) Effect of $k_{root}$ on the photosynthesis (mid-bottom) hourly variation of stomatal conductance and leaf water potential for the designed ideotype (red) and modified ideotype (blue).

**Table 1** Management/Agronomic parameters, plant physiological parameters of typical maize, and bio-chemistry parameters for gas exchange calculations.

| Management/Agronomic parameters | |
|---|---|
| $b$, Soil-texture-dependent parameter (unit less) | 14.95 |
| $k_{sat}$, Saturated soil hydraulic conductivity [mol m$^{-1}$ s$^{-1}$ MPa$^{-1}$] | 1.69 |
| $\psi_{sat}$ [MPa] | -0.00598 |
| $\theta_{sat}$ (Saturated water content in the soil) [m$^3$ m$^{-3}$] | 0.39 |
| $H_s$ (depth of soil) [m] | 1 |
| Rs, Radius of soil occupied/supplied for one plant [m] | 0.1128 |
| Irrigation frequency (day) | 7 |
| **Plant physiological parameters** | |
| L, Root length density (m m$^{-3}$) | 15200 |
| r$_{root}$, Root radius at the end of rhizosphere (m) | 0.0005 |
| d, leaf width (m) | 0.1 |
| H$_c$, Height of the plant (m) | 1 |
| Permanent wilting soil water potential for leaf wilting (MPa) | -1.33 |
| Temperature for permanent leaf damage ($^0$C) | 60 |
| LAI, Leaf area index, [m$^2$m$^{-2}$] | 1 |
| **Plant Hydraulic Parameters** | |
| $g_{min}$, minimum stomatal conductance mol m$^{-2}$s$^{-1}$ | 0.02 |
| $g_{max}$, maximum stomatal conductance mol m$^{-2}$s$^{-1}$ | 0.25 |
| $S_l$, response of stomatal conductance with leaf water potential mol m$^{-2}$s$^{-1}$MPa$^{-1}$ | 15 |
| $S_r$, response of stomatal conductance upon sensing drying soil mol m$^{-2}$s$^{-1}$MPa$^{-2}$ | -1.25 |
| $\psi_{th}$, threshold of leaf water potential for stomatal closure, MPa | 200 |
| $K_{leaf}$, hydraulic conductance of leaf mmol m$^{-2}$s$^{-1}$MPa$^{-1}$ | 45 |
| K$_{root}$, hydraulic conductance of root mmol m$^{-2}$s$^{-1}$MPa$^{-1}$ | 45 |
| K$_{root}$, hydraulic conductance of stem mmol m$^{-2}$s$^{-1}$MPa$^{-1}$ | 45 |

| Bio-chemistry parameters from gas exchange data | | | | | |
|---|---|---|---|---|---|
| $g_1$ (unitless) | 0.0036 | $J_b$ (unitless) | 4.93 | $Q_{10,V_{PEP\max}}$ (unitless) | 1.39 |
| $g_2$ (unitless) | 1.1693 | x (unitless) | 0.0844 | $R_{d,Q10}$ (unit less) | 1.55 |
| $Q_{10,V_{PEP},R_{\max}}$ (unit less) | 13.1 | $\alpha$ (unitless) | 0.00445 | $J_a$ (unit less) | 5.57 |
| $Q_{10,V_{RO\max}}$ (unit less) | 0.945 | $\beta$ (unitless) | 0.0715 | $R_{t,25}$ [mol m$^{-2}$ s$^{-1}$] | 2.28 |
| A (unit less) | 0.094 | $J_{\max,25}$ [mol m$^{-2}$ s$^{-1}$] | 322.3 | $V_{PEP,R,25}$ [mol m$^{-2}$ s$^{-1}$] | 96.1 |
| B (unit less) | 31.6 | $V_{PEP\max,25}$ [mol m$^{-2}$ s$^{-1}$] | 51 | | |

| C (unit less) | 42.1 | $V_{RO_{max,25}}$ [mol m$^{-2}$ s$^{-1}$] | 126.7 | | |

**Table 2** Bounds to hydraulic parameters varied in the genetic algorithm.

| | $g_{min}$ | $g_{max}$ | $S_l = \tan\theta_l$ | $S_r = \tan\theta_r$ | $\psi_{th}$ | $K_{leaf+stem}$ | $K_{root}$ |
|---|---|---|---|---|---|---|---|
| | mol m$^{-2}$s$^{-1}$ | | mol m$^{-2}$s$^{-1}$MPa$^{-1}$ | mol m$^{-2}$s$^{-1}$MPa$^{-2}$ | MPa | mmol m$^{-2}$s$^{-1}$ MPa$^{-1}$ | |
| **Lower bound** | 1e-6 | 0.07 | $\theta_l = 0$ | $\theta_r = 0$ | -1.33 | 1e-6 | 1e-6 |
| **Upper bound** | 3 | 3.0 | $\theta_l = 89.99$ | $\theta_r = 89.99$ | 0 | 30 | 60 |

**Nomenclature**

| | | |
|---|---|---|
| $a$ | - | Absorptance of leaf |
| $a_{IR}$ | - | Absorptance of leaf for thermal infrared radiation |
| $A$ | - | Constant related with carboxylation rate |
| $An$ | [mol m$^{-2}$ s$^{-1}$] | Hourly net assimilation or net photosynthesis |
| $b$ | - | Soil texture dependent parameter |
| $b_0$ | - | Parameter related with probabilistic GA formulation |
| $b_1$ | - | Parameter related with probabilistic GA formulation |
| $b_2$ | - | Parameter related with probabilistic GA formulation |
| $B$ | - | Constant related with carboxylation rate |
| $c_0$ | kPa | Constant related with Tetens formula |
| $c_1$ | - | Constant related with Tetens formula |
| $c_2$ | °C | Constant related with Tetens formula |
| $C$ | - | Constant related with carboxylation rate |
| $C_{c,a}$ | [Pa Pa$^{-1}$] or ppm | Atmosphere CO$_2$ partial pressure, or concentration |
| $C_{c,i}$ | [Pa Pa$^{-1}$] or ppm | Intercellular airspace CO$_2$ partial pressure, or concentration |
| $C_{c,m}$ | [Pa Pa$^{-1}$] | Mesophyll CO$_2$ partial pressure, or concentration. |
| $C_P$ | [J mol$^{-1}$K$^{-1}$] | Specific heat of air. |
| $d$ | [m] | Leaf average width |
| $d_e$ | [m] | Parameter related with leaf width |
| $D_H$ | [m$^2$s$^{-1}$] | Thermal diffusivity in air |
| $D_{WV}$ | [m$^2$s$^{-1}$] | Water vapor diffusivity in air |
| $f_{PAR\_PSII}$ | - | Fraction of PAR contributes to the Photosystem II |
| $g_1$ | - | lant physiological parameter related with photosynthesis or $CO_2$ assimilation |
| $g_2$ | - | lant physiological parameter related with photosynthesis or $CO_2$ assimilation |
| $g_{blc}$ | [mol m$^{-2}$ s$^{-1}$] | Boundary conductance to water transport |
| $g_{hbc}$ | [mol m$^{-2}$ s$^{-1}$] | Boundary conductance to heat transfer on leaf surface. |
| $g_{min}$ | [mol m$^{-2}$ s$^{-1}$] | Minimum Stomatal conductance, or stomatal conductance at light compensation point, minimum stomatal conductance to water vapor including epidermal conductance |
| $g_{max}$ | [mol m$^{-2}$ s$^{-1}$] | Maximum Stomata conductance |
| $g_{C,bs}$ | [mol m$^{-2}$ s$^{-1}$] | Bundle-sheath conductance to $CO_2$ |
| $g_{st}$ | [mol m$^{-2}$ s$^{-1}$] | Effective stomata conductance to water |
| $Gr$ | - | Grashof number |
| $H_s$ | [m] | Depth of soil |
| $H_c$ | [m] | Height of the plant |
| $H_m$ | [m] | Height at which wind speed obtained from weather data |
| $I$ | [mol m$^{-2}$ s$^{-1}$] | A parameter related with electron transport rate |
| $J_a$ | [mol m$^{-2}$ s$^{-1}$] | Physiological parameter related with electron transport rate |
| $J_b$ | [mol m$^{-2}$ s$^{-1}$] | Physiological parameter related with electron transport rate |
| $J_{C,d}(enzyme)$ | [mol m$^{-2}$ s$^{-1}$] | $CO_2$ demand by photosynthetic activity based enzyme limited condition |
| $J_{C,d}(light)$ | [mol m$^{-2}$ s$^{-1}$] | $CO_2$ demand by photosynthetic activity based sunlight limited condition |
| $J_{e,t}$ | [mol m$^{-2}$ s$^{-1}$] | total electron transport rate is at leaf temperature |
| $J_{e,25}$ | [mol m$^{-2}$ s$^{-1}$] | total electron transport rate is at 25 ° C |
| $J_{emax,25}$ | [mol m$^{-2}$ s$^{-1}$] | Maximum total electron transport rate is at 25 ° C |

| Symbol | Units | Description |
| --- | --- | --- |
| $J_{W,d}$ | [mol m$^{-2}$ s$^{-1}$] | Rate of water vapor demanded by atmosphere from leaf |
| $J_{W,s}$ | [mol m$^{-2}$ s$^{-1}$] | Rate of water supplied from soil to leaf |
| $k_{sat}$ | [mol m$^{-2}$ s$^{-1}$ Pa$^{-1}$] | hydraulic conductivity of saturated soil |
| $K_{leaf}$ | [mol m$^{-2}$ s$^{-1}$ Pa$^{-1}$] | leaf hydraulic conductance to water |
| $K_p$ | [mol m$^{-2}$ s$^{-1}$ Pa$^{-1}$] | the Michaelis-Menten constant |
| $K_{plant}$ | [mol m$^{-2}$ s$^{-1}$ Pa$^{-1}$] | plant hydraulic conductance to water |
| $K_{root}$ | [mol m$^{-2}$ s$^{-1}$ Pa$^{-1}$] | root hydraulic conductance to water |
| $K_{soil}$ | [mol m$^{-2}$ s$^{-1}$ Pa$^{-1}$] | soil hydraulic conductance to water |
| $K_{stem}$ | [mol m$^{-2}$ s$^{-1}$ Pa$^{-1}$] | stem hydraulic conductance to water |
| $L$ | [m m$^{-3}$] | Root length density, root length per unit volume of soil |
| $L_{vap}$ | [J mol$^{-1}$] | Latent heat of vaporization of water. |
| $LAI$ | [m$^2$ m$^{-2}$] | Leaf area index |
| $m$ | - | a factor related with zero plane displacement for wind speed |
| $n$ | - | a factor related to the momentum roughness parameter for wind speed |
| $photo$ | [mol m$^{-2}$ s$^{-1}$] | Hourly net CO2 assimilation |
| $Photo$ | [mol m$^{-2}$] | Total photo |
| $PAR$ | [mol m$^{-2}$ s$^{-1}$] | Photo active radiation |
| $P_o$ | - | Initial population in genetic algorithm |
| $P_a$ | [Pa] | Atmospheric pressure |
| $Pr$ | | Prandtl number |
| $P_{va}$ | [Pa] | Vapor pressure of air |
| $P_{vl}$ | [Pa] | Vapor pressure of leaf surface |
| $Q_{10,Rt}$ | - | Q10 coefficient conversion factor related to mitochondrial respiration calculation |
| $Q_{10,V_{PEP\max}}$ | - | Q10 coefficient conversion factor related to maximum PEP carboxylation |
| $Q_{10,V_{PEP,R\max}}$ | - | Q10 coefficient conversion factor related to maximum PEP regeneration |
| $Q_{10,V_{RO\max}}$ | - | Q10 coefficient conversion factor related to maximum rubisco carboxylation |
| $r$ | - | Reflectance, i.e. amount of sunlight reflected from the surroundings, |
| $r_{root}$ | [m] | Root radius including rhizosphere |
| $R_t$ | [mol m$^{-2}$ s$^{-1}$] | Total mitochondrial respiration in the mesophyll and bundle sheath at leaf temperature |
| $R_{t,25}$ | [mol m$^{-2}$ s$^{-1}$] | Total mitochondrial respiration in the mesophyll and bundle sheath at 25° C temperature |
| Re | - | Reynolds number |
| $RH$ | - | Relative humidity of surrounding air. |
| $S$ | [W m$^{-2}$] | Solar radiation, |
| $S_c$ | - | Schmidt number |
| $S_l$ | [mol m$^{-2}$ s$^{-1}$ MPa$^{-1}$] | Slope of stomatal conductance with leaf water potential |
| $S_r$ | [mol m$^{-2}$ s$^{-1}$ MPa$^{-2}$] | Slope of $S_l$ with root water potential |
| $T_a$ | [° C] | Temperature of air |
| $T_{leaf}$ | [° C] | Temperature of leaf |
| $T_{wilt}$ | [° C] | leaf temperature at permanent leaf damage |
| $Tr$ | [mol m$^{-2}$ s$^{-1}$] | Rate of water transpires from soil to environment through the plant |
| $U_c$ | [m s$^{-1}$] | Wind speed on the canopy |
| $U_m$ | [m s$^{-1}$] | Wind speed from weather data |
| $U^*$ | [m s$^{-1}$] | Shearing velocity |
| $V_{PEP}$ | [mol m$^{-2}$ s$^{-1}$] | Effective Rate of PEP carboxylation at leaf temperature given by Michaelis-Menten Equation |
| $V_{PEP_{max}}$ | [mol m$^{-2}$ s$^{-1}$] | Maximum PEP carboxylation rate at leaf temperature |
| $V_{PEP_{max,25}}$ | [mol m$^{-2}$ s$^{-1}$] | Maximum PEP carboxylation rate at 25 C temperature |
| $V_{PEP,R}$ | [mol m$^{-2}$ s$^{-1}$] | PEP regeneration rate at leaf temperature |
| $V_{PEP,R,25}$ | [mol m$^{-2}$ s$^{-1}$] | PEP regeneration rate at 25 C temperature |
| $V_{RO_{max}}$ | [mol m$^{-2}$ s$^{-1}$] | Maximum rubisco carboxylation rate at leaf temperature |
| $V_{RO_{max,25}}$ | [mol m$^{-2}$ s$^{-1}$] | Maximum rubisco carboxylation rate at 25 C temperature |
| $VPD$ | [Pa Pa$^{-1}$] | Vapor pressure deficit between intercellular space and atmosphere |
| Water Loss | [mol m$^{-2}$ s$^{-1}$] | Transpiration of water through the plant |
| $x$ | - | is a fitting parameter related to photosynthesis rate |
| $Z$ | - | parameter to make the exponent dimensionless in stomatal conductance model |

**Greek**

| | | |
|---|---|---|
| $\alpha$ | - | empirical parameter related with boundary layer conductance |
| $\beta$ | - | ratio of $CO_2$ conductance and water vapor conductance through stomata |
| $\lambda$ | - | Empirical curvature factor related with electron transport rate |
| $\chi$ | - | ratio of $CO_2$ conductance and water vapor conductance through air boundary layer |
| $\nu_a$ | [m$^2$ s$^{-1}$] | kinetic viscosity of air |
| $\theta$ | [m$^3$ m$^{-3}$] | Volumetric water content in the soil |
| $\theta_{sat}$ | [m$^3$ m$^{-3}$] | Soil volumetric saturation |

| Symbol | Units | Description |
|---|---|---|
| $\psi_{leaf}$ | [Pa] | Leaf water potential |
| $\psi_{root}$ | [Pa] | Root water potential |
| $\psi_{sat}$ | [Pa] | Saturated soil water potential |
| $\psi_{soil}$ | [Pa] | Soil water potential |
| $\psi_{soil,root}$ | [Pa] | Soil water potential at root |
| $\psi_{th}$ | [Pa] | Threshold leaf water potential at stomatal closure |
| $\sigma$ | [W m$^{-2}$ K$^{-4}$] | Stefan Boltzmann constant |

# Supplementary Materials

# Integrating optimization with thermodynamics and plant physiology for crop ideotype design

**Talukder Z. Jubery[1], Baskar Ganapathysubramanian[1], Matthew E. Gilbert[2], and Daniel Attinger[1]**

[1]Department of Mechanical Engineering and Department of Electrical and Computer Engineering, Iowa State University, Ames, IA, 50011;

[2]Department of Plant Sciences, University of California, Davis, CA 95616, USA.
*Correspondence to: *baskarg@iastate.edu, gilbert@ucdavis.edu, attinger@iastate.edu*

**Justification for selecting June-July weather**

Maximum solar radiance is observed in this period of year in the northern hemisphere. Our hypothesis was that a crop should be the most productive in this period provided that the plant has access to adequate water and nutrients. Due to high solar radiance that results high temperature, and generally low relative humidity in drought-prone areas, this period should also mimic the highest water demand by the environment from the plant.

We considered that the typical maize plant was fully grown, i.e. at the beginning of full canopy closure, and total yielding period was two months. During this two-month period average solar radiation was 489.17 W/m2, relative humidity 45.63%, air temperature was 30.6 $^0$C, and wind speed was 3.048 m/s, precipitation does not occur in drought conditions.. The hourly variations of weather parameters are in Figure S1.

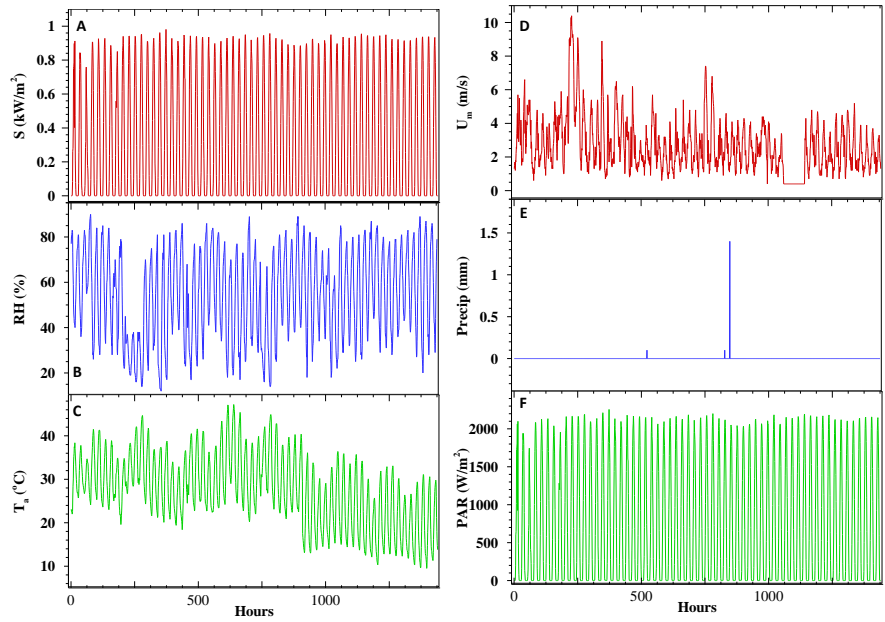

**Fig. S1** Hourly variation of solar radiation (S), relative humidity (RH), atmospheric temperature ($T_a$), wind speed ($U_m$), precipitation (Precip) and photosynthetically active radiation (PAR) in June-July 2010, Davis, CA.

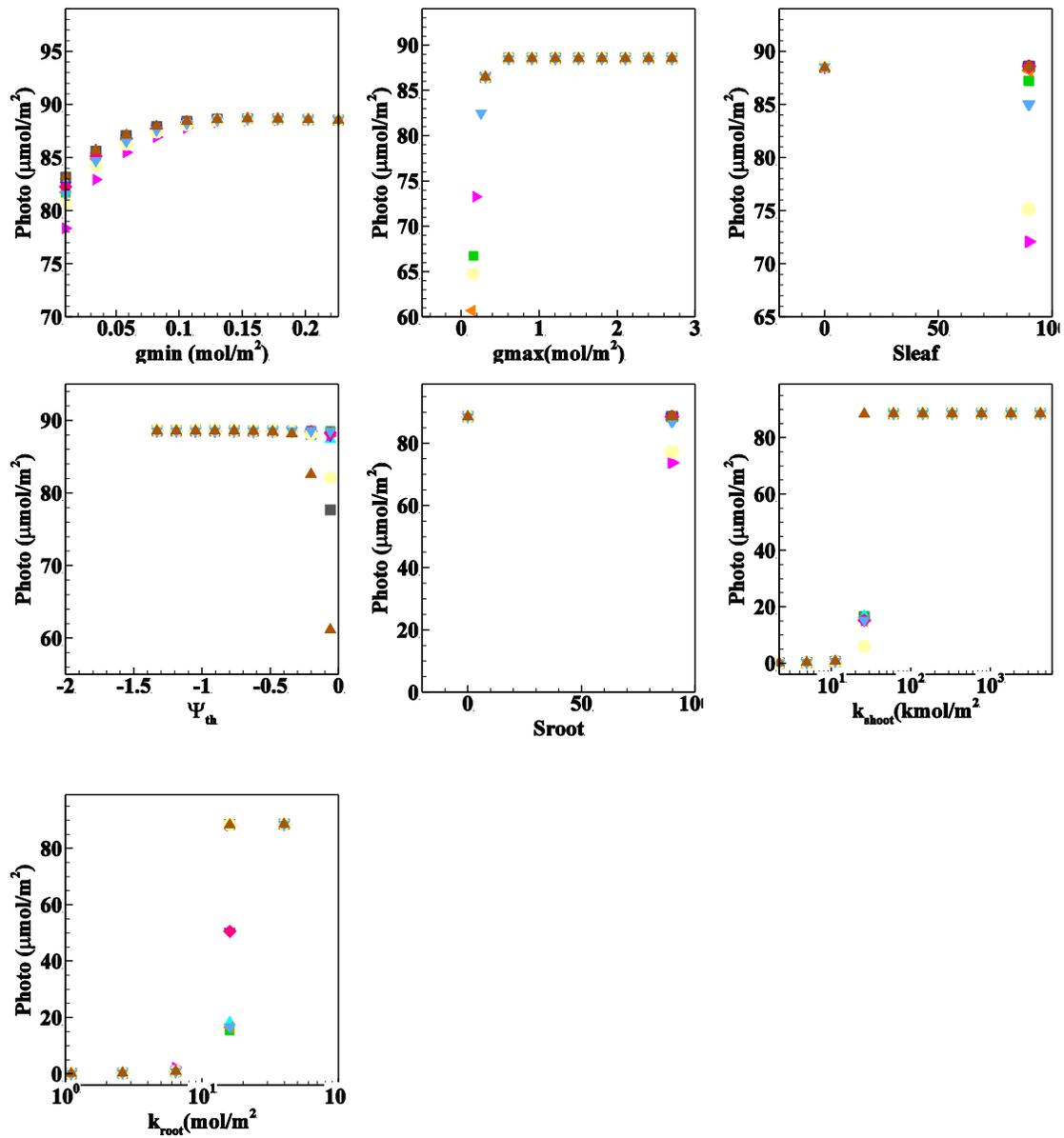

**Figure S2:** Sensitivity analysis of the traits obtained from GA optimization.